\documentclass[a4paper,fleqn,usenatbib,useAMS,usefloat,referee ]{mnras}
\usepackage[normalem]{ulem}
\usepackage[T1]{fontenc}
\usepackage{ae,aecompl}
\usepackage{longtable}

\usepackage[graphicx]{realboxes}
\usepackage{epstopdf}
\usepackage{xcolor}
\usepackage{amsmath}	% Advanced maths commands
\usepackage{amssymb}	% Extra maths symbols
\usepackage{subfig}
\usepackage{ulem}
\usepackage{booktabs}
\usepackage{float}
\usepackage{array}
\usepackage{caption}
\usepackage{scalerel}
\usepackage{tabularx}
\usepackage{hyperref}
\usepackage{footmisc}
\hypersetup{
    colorlinks=true,
    linkcolor=blue,
}
\title[Dormant Black Holes in OGLE Galactic Ellipsoidals]
{Search for Dormant Black Holes in Ellipsoidal Variables III. The OGLE BULGE Short-Period Sample}
\author[Gomel, Faigler, Mazeh and Pawlak]
{Roy Gomel,$^{1}$\thanks{E-mail:
roygomel@tauex.tau.ac.il}
Simchon Faigler,$^{1}$
Tsevi Mazeh$^{1}$
and
Micha{\l} Pawlak$^{2}$
\\
$^{1}$School of Physics and Astronomy, %Raymond and Beverly Sackler Faculty of Exact Sciences, \\
Tel Aviv University, Tel Aviv, 6997801, Israel\\
$^{2}$Astronomical Observatory, Jagiellonian University, ul. Orla 171, 30-244 Krak{\'o}w, Poland\\
}

\date{Accepted XXX. Received YYY; in original form ZZZ}

\begin{document}
\label{firstpage}
\pagerange{\pageref{firstpage}--\pageref{lastpage}}
\maketitle
%
%======================================
\begin{abstract}

This is the third of a series of papers that presents an algorithm to search for close binaries with massive, possibly compact, unseen secondaries.
The detection of such a binary is based on identifying a star that displays a large ellipsoidal periodic modulation, induced by tidal interaction with its companion.
%
%Based on the observed ellipsoidal amplitude and the primary mass and radius, one can derive a minimum mass ratio of the binary. 
%defined as the mass ratio obtained for an inclination of $90^{\circ}$.
%
%A binary with a minimum mass ratio significantly larger than unity might be a candidate for having a dormant compact-object companion. 
%
%Unfortunately, in many cases the primary mass and radius are not well known. Therefore, we presented
%
In the second paper of the series we presented a simple approach to derive 
%that circumvents this problem by suggesting 
a robust {\it modified} minimum mass ratio (mMMR),
based on the observed ellipsoidal amplitude, without knowing the primary mass and radius,
assuming the primary fills its Roche lobe. 
The newly defined mMMR is always {\it smaller} than the actual mass ratio. Therefore, a binary with an mMMR larger than unity is a good candidate for having a massive secondary, which might be a black hole or a neutron star. 
This paper considers $10,956$ OGLE short-period ellipsoidals observed towards the Galactic Bulge.
We re-analyse their modulation and identify $136$ main-sequence systems with mMMR significantly larger than unity as candidates for having compact-object secondaries, assuming their observed periodic modulations reflect indeed the ellipsoidal effect.  Obviously, one needs follow-up observations to find out the true nature of these companions.
\end{abstract}
%===================================

%

%===========================
\section{Introduction}
\label{sec:intro}
%===========================

This is the third of a series of papers (Paper I: \citealp*{gomel21a}, Paper II: \citealp*{gomel21b}) 
that presents an algorithm to search large sets of photometric light curves for evidence of close binaries with dormant black holes (BH),
and, in some cases, dormant neutron stars (NS). Unlike known low-mass BH binaries in the Galaxy \citep{corral16,tetarenko16}, dormant BH/NS binaries do not emit x-ray since their primaries do not transfer mass onto the compact object, or they are in a quiescent stage.
The detection of such a binary is based on identifying a star that displays a large ellipsoidal periodic modulation, induced by tidal interaction with its companion.

Based on the observed ellipsoidal amplitude and the estimated mass and radius of the primary star, one can derive a minimum mass ratio of the binary, 
defined as the mass ratio obtained for an inclination of $90^{\circ}$,
 provided most of the light is coming from the primary star \citep[e.g.,][]{faigler11,faigler15}.
A binary with a minimum mass ratio significantly larger than unity might be a candidate for having a dormant compact-object companion. 

Unfortunately, in many cases the primary mass and radius are not well known.
Therefore, \hyperlink{cite.gomel21b}{Paper II}
presented a simple approach that circumvents this problem by suggesting a robust {\it modified} minimum mass ratio (mMMR), assuming the primary fills its Roche lobe. 
The newly defined mMMR depends on the amplitude of the second harmonic of the modulation, and to some extent on the primary temperature, but does not depend on the mass or radius of the primary star. 

The mMMR is always {\it smaller} than the minimum mass ratio, which is, in its turn, smaller than the actual mass ratio. Therefore, binaries with a  modified minimum mass ratio significantly larger than unity are good candidates for having a compact object secondary,
even though we cannot reliably constrain their primary mass and radius.

In this paper, we consider the OGLE collection of $25,405$ ellipsoidal binary systems in the Galactic bulge \citep{soszy16}. 
As the paper shows below, the OGLE ellipsoidals can be divided into short-, medium- and long-period systems, which we interpret as systems with main-sequence,  red-clump, and red-giant primaries, respectively.

Our analysis is focused on the short-period binaries. In case the primary is not a main-sequence (MS) star, a large mass ratio is not necessarily an indication of a compact object. 
As pointed out by \hyperlink{cite.gomel21b}{Paper II}, 
Algol-type binaries, with sub-giant or giant primaries, are famous counter examples. These systems, which probably went through a mass-transfer phase during their evolution \citep[e.g.,][]{algol18, chen20}, can have a mass ratio larger than unity and still have an MS secondary \citep[e.g.,][]{negu18,samadi18}.

%{\color{red}These systems,  that converted their primordial mass ratios, 
%The recently suggested systems consisting of an evolved primary and a dormant compact-object secondary \citep[e.g.,][]{thompson19, liu19, rivinius20, jayasinghe21}, could be Algol-type binaries \citep[e.g.,][]{van-den-Heuvel20, irrgang20, shenar20, bodensteiner20, mazeh20, el-badry20}. 

This work uses the publicly available OGLE $I$- and $V$-band light curves of the OGLE Galactic-bulge ellipsoidals \citep{soszy16} and the derived OGLE period to re-analyze the periodic modulation and obtain the $I$-band amplitudes of the first four harmonics of the orbital period. We then derive the mMMR of each system, based on the amplitude of the second harmonic, identifying $136$ short-period binaries as candidates 
%with lower mMMR > $1$
%percentile 
%$\hat{q}_{\rm min}^{-1\sigma} , 
that might have compact-object secondaries,
provided these systems are indeed ellipsoidal variables.
Throughout the diagrams of the paper we denote these candidates with green, so the reader can identify them in all figures.

Section~\ref{sec:selection} studies the general features of the OGLE ellipsoidals and the separation between the short-period variables and the rest of the sample, 
Section~\ref{sec:analysis} details our re-analysis of the short-period ellipsoidals, and
Section~\ref{sec:candidates} presents the sample of candidates that might have secondary massive unseen companions, including a discussion of OGLE BLG-ELL-024717 --- a typical example of a binary that might have a BH secondary. Finally. Section~\ref{sec:discussion} discusses and summarises our findings.

%================================
\section{OGLE GALACTIC-BULGE ELLIPSOIDALS Sample}
\label{sec:selection}
%=================================

Our analysis is based on the OGLE-IV catalog of Bulge ellipsoidals  \citep{soszy16}
accessible through the public website of the OGLE Collection of Variable Stars.\footnote{http://ogledb.astrouw.edu.pl/$\sim$ogle/OCVS/} 
%The identifiers OGLE-BLG-ELL-NNNNNN (where NNNNNN is a six-digit number) have been given to each ellipsoidal binary in the catalog, and the stars are arranged in order of increasing right ascension. 
Each of the objects in the catalog has been visually verified by the OGLE team, therefore the level of contamination of the sample is expected to be low. For technical details of the OGLE survey refer to \citet{udalski15}.

\begin{figure} 
\centering
{  \includegraphics[scale=0.75]{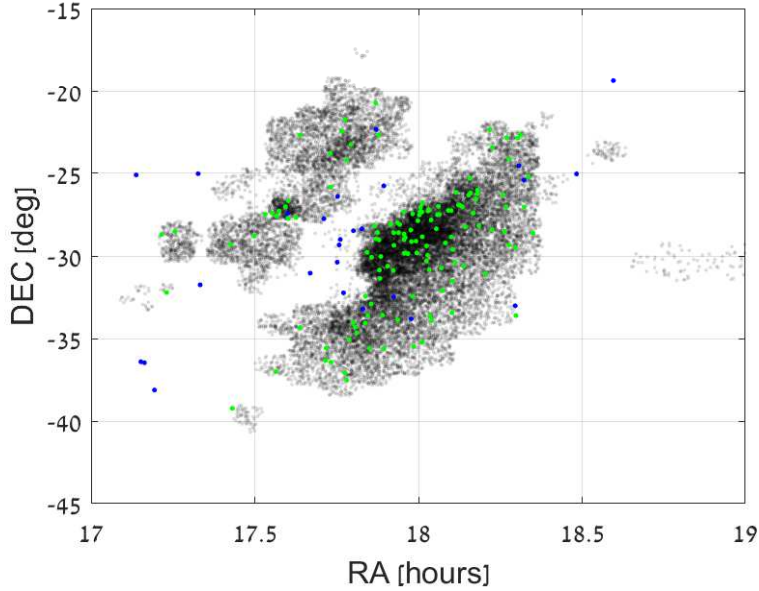}  }
\caption{Spatial distribution of $25,405$ OGLE Bulge ellipsoidal binary systems towards the Galactic bulge. Candidates with mMMR significantly larger than unity are colored in green. 26 out of 59 Galactic BHBs around the OGLE field are shown by blue points.
}
\label{fig:Coo}
\end{figure}

The spatial distribution of all ${25,405}$ OGLE Bulge ellipsoidal variables is presented in Fig.~\ref{fig:Coo}. 
The figure shows in green the sky locations of stars with mMMR significantly larger than unity, identified as candidates for having compact-object secondaries 
%The way in which we selected the candidates is described in detail in .
(see Section~\ref{sec:candidates}).
One can see that the candidates are spread over the whole OGLE observed regions, with no preferred location.

%--------------------------------
%----------------------------------------

It is well known that most known black-hole binaries (BHBs) are concentrated towards the direction of the Galactic bulge \citep[e.g.,][]{corral16}. Thus, we also plotted in Fig.~\ref{fig:Coo} the position of 26 out of the 59 known Galactic BHBs which are located around the OGLE fields. Unfortunately, none of these systems have available photometry in the public database of the OGLE project.

\begin{figure} 
\centering
{\includegraphics[scale=0.75]{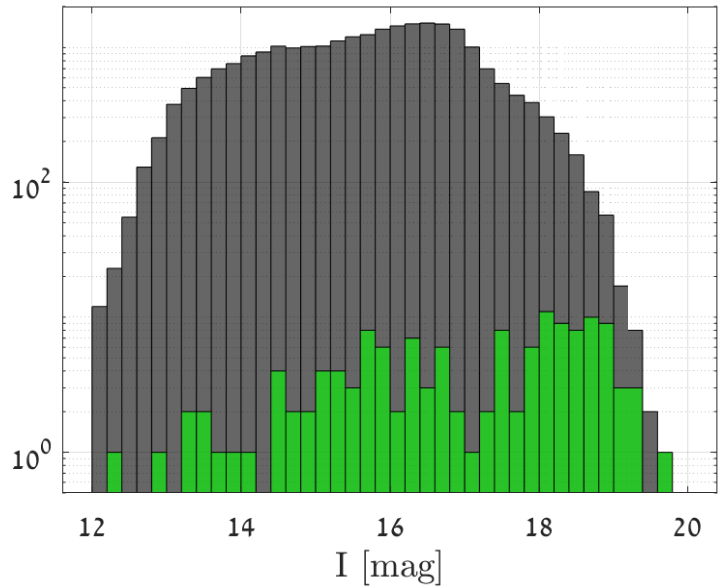}}
\caption{ Distribution of $I$-band magnitude of $25,405$ OGLE Bulge ellipsoidal binary systems (black) 
and of candidates with mMMR significantly larger than unity (green). 
}
\label{fig:Imag-hist}
\end{figure}

Fig.~\ref{fig:Imag-hist} shows an OGLE $I$-mag histogram of the ellipsoidals. This is the magnitude at maximum light, derived by a light-curve template fitting \citep{soszy16}. Note that the $I$ mag given in Table~\ref{tab:Data} is slightly different and represents the mean magnitude of those ellipsoidals. 
The figure suggests that the OGLE sample has high completeness up to $I \sim 17$ mag, but  includes systems up to $I \sim 20$ mag.
The figure shows also the brightness distribution of the compact-object candidates, which seems as if skewed towards the faint end of the sample. This is so because we chose the compact-secondary candidates from the short-period ellipsoidals, which are at the faint end of the OGLE ellipsoidals, as is shown in Fig.~\ref{fig:P-WI}. 

%----------------------------------------
%------------------------------------------

\newpage
%-----------------------------------------------------------
\subsection{Dividing the ellipsoidals into three sub-samples: main-sequence, sub-giant, and giant primaries}
\label{subsection:three_groups}
%-----------------------------------------------------------

\begin{figure} 
\centering
{  \includegraphics[scale=0.75]{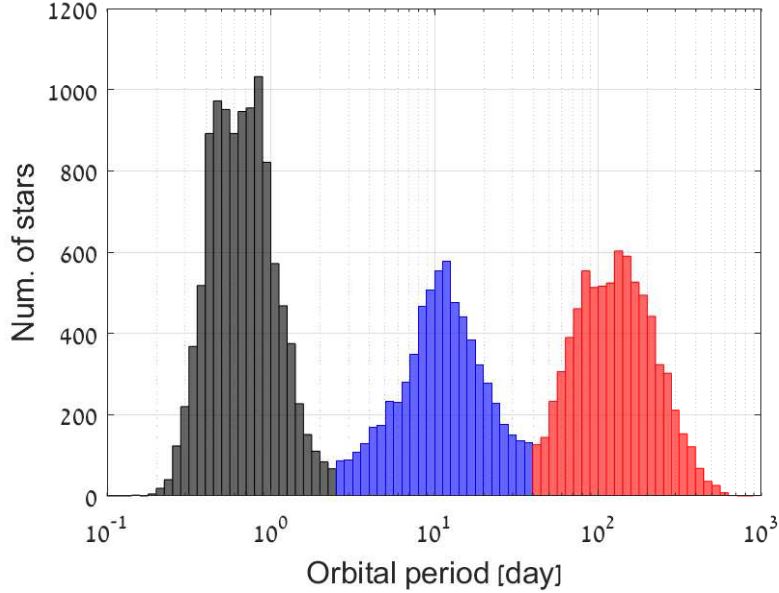}  }
\caption{Period distribution of OGLE Bulge ellipsoidals, classified into three groups: short-period (black), mid-period (blue) and long-period (red). The distribution is plotted with equally-spaced bins of $\Delta \log P = 0.05$.
}
\label{fig:P-hist}
\end{figure}

Fig.~\ref{fig:P-hist} presents a period histogram of the entire OGLE ellipsoidal Bulge sample. 
The histogram suggests that the OGLE ellipsoidals consist of three distinct groups: $10,956$ short- (P $\leq$ $2.5$ days), $6,751$ mid- ($2.5$ days < P $\leq$ $40$ days) and $7,698$ long-period (P > $40$ days) binaries.
%(see also Fig.~\ref{fig:P-Ri2}).

%--------------------------------------
%----------------------------------------

\begin{figure} 
\centering
{  \includegraphics[scale=0.7]{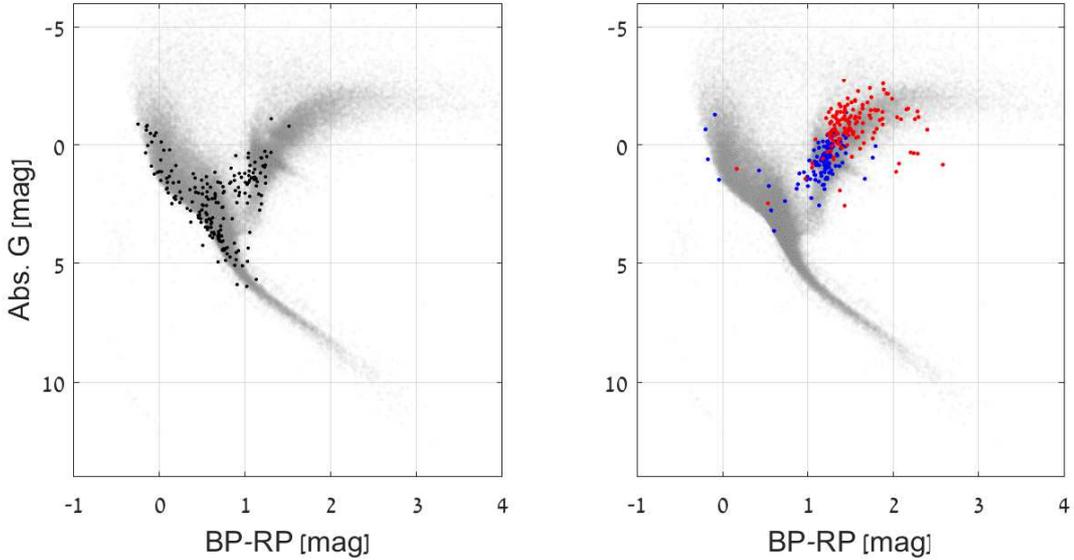}  }
\caption{OGLE Bulge ellipsoidals on Gaia CMD. Only systems with parallax-over-error > 5, and available extinction-correction from GAIA are plotted. Systems are classified into three groups: 195 short-period (black), 122 mid-period (blue) and 138 long-period (red) binaries. As a background, a grey-scale density map of Hipparcos stars is plotted, used as a proxy for the expected CMD in the solar neighbourhood, as done, for example, by \citet{SM19}.
}
\label{fig:CMD_Gaia}
\end{figure}

To further explore the division between the three sub-samples of the OGLE ellipsoidals we plot in Fig.~\ref{fig:CMD_Gaia} some of these systems on the Gaia color-magnitude diagram (CMD). 
We include only stars with parallax-over-error > 5 and available extinction correction from Gaia. Note that the position of the stars on the diagram is susceptible to errors of the extinction correction that may be highly inaccurate \citep{GaiaDR2,andrae18,anders19}. 

%-----------
%--------------------------------------------------

The figure suggests that short-period ellipsoidals consist mainly of systems with main-sequence (MS) primaries, mid-period systems occupy the red-clump (RC) region, and long-period binaries lay on the Red-Giant Branch (RGB).
The association of the short-period ellipsoidals with MS primaries is due to the fact that 
the ellipsoidal amplitude depends on the ratio of the primary radius to the orbital separation to the third power. Therefore, binaries with main-sequence primaries can reveal their ellipsoidal modulation only if they reside in short-period systems.

The small sample of Fig.~\ref{fig:CMD_Gaia} suggests that about a fourth of the short-period ellipsoidals contain a sub-giant primary. This is consistent with the overlap between the short- and mid- period distribution seen in Fig.~\ref{fig:P-hist}. We include these short-period ellipsoidals in our analysis, as it should be relatively simple to determine whether their companion is a compact object in case the derived mass ratio is larger than unity. This is so because sub-giant primaries in short-period binaries are only slightly evolved, and therefore cannot easily obscure a more massive main-sequence secondary. 

\begin{figure} 
\centering
{  \includegraphics[scale=0.75]{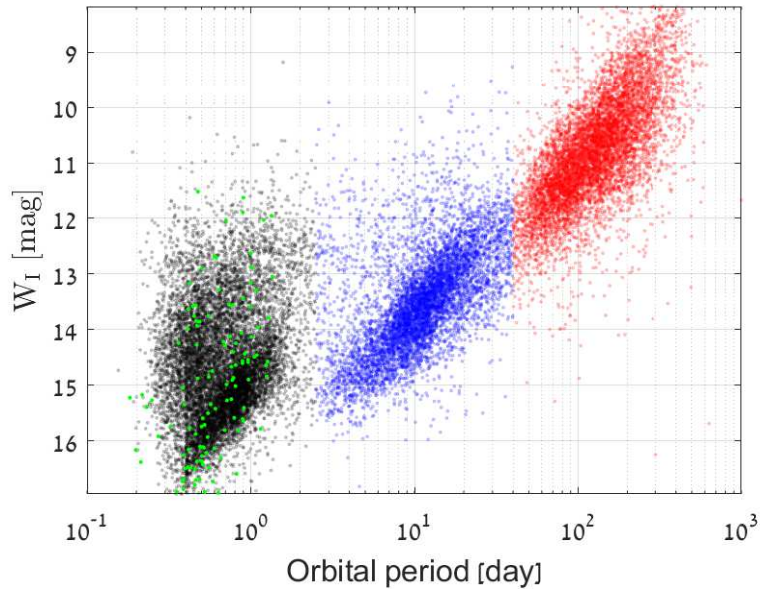}  }
\caption{ Wesenheit index $W_{\rm I}$, as a function of orbital period, classified into three groups: $10,956$ short-period (black ), $6,751$ mid-period (blue) and $7,698$ long-period (red) ellipsoidal variables. Candidates with mMMR significantly larger than unity are colored in green.  Data within Y-axis $0.1$--$99.9$ percentiles are shown. 
}
\label{fig:P-WI}
\end{figure}

Another aspect of dividing the Bulge ellipsoidals into three sub-samples can be seen in Fig.~\ref{fig:P-WI}, which shows a period-luminosity (PL) diagram of the MS, RC, and RGB ellipsoidals. For this purpose, the Wesenheit index, $W_{\rm I} = I - 1.14 (V-I)$, was calculated using OGLE $I$- and $V$-band magnitudes at maximum light, with the Bulge calibration developed by \citet{pietrukowicz2015}. Here also there is a clear division between MS and evolved stars which form two clearly different PL relations. These relations will be discussed below.

%--------------------------------------
%----------------------------------------

\newpage
%================================
\section{Analysis}
\label{sec:analysis}
%=================================

As explained in the introduction, this paper is focused on the short-period OGLE ellipsoidals, assuming most of these systems' primaries are on the main sequence.

Our analysis is based primarily on the OGLE-IV
$I$-band photometry, and for 234 systems, when it was not available, on OGLE-II or OGLE-III data.\footnote{http://ogledb.astrouw.edu.pl/$\sim$ogle/OCVS/}
No systematic effects are expected in our analysis as the OGLE data from all four phases are calibrated to the Landoldt photometric standard system \citep{landolt92}.
The OGLE-IV data include six years of observations from 2010 to 2015. The number of points varies from about 100 to more than 10,000 epochs per light curve due to different OGLE observation strategies in different parts of the Bulge. 

In our analysis, each light curve was divided into 1-year seasons. Within each season we removed data points that deviated more than $10\sigma$ from the median value of the season, where $\sigma$ was calculated as $1.48$ times the Median-Absolute-Deviation (MAD) of that season's light curve. 
Data were binned into equally-spaced bins of 10 minutes;  
keeping original time and magnitude of the OGLE measurement for bins with a single point, while replacing the measurements with the weighted-average of magnitudes and times for bins with multiple points.
We excluded seasons with less than 15 points and light curves with less than 100 points. Starting with a short-period sample of $10,956$, we ended up with $10,814$ systems.

We used the OGLE period for fitting a four-harmonic model to the observed magnitudes:
\begin{equation}
A_0 + \sum_{i=1}^4 a_{\rm ic} \cos\big(\frac{2 \pi i}{P} (t-T_0)\big) + a_{\rm is} \sin\big(\frac{2 \pi i}{P} (t-T_0)\big),
\label{eq:harmonics}
\end{equation}
where P is the OGLE period, $a_{\rm ic}$ ($a_{\rm is}$) is the cosine (sine) coefficient of the $i$-th harmonic, $A_0$ is the stellar averaged magnitude, fitted for each season separately, and $T_0$ is chosen such that $a_{\rm 2s}=0$. 
We performed a robust regression fit --- an algorithm that iteratively minimizes the weighted residuals least-squares, and is less sensitive to outliers \citep{leroy87}.

The amplitude of each harmonic is simply
%---------------------------------------------------------------
\begin{equation}
A_i = \ 
\sqrt{a^2_{ic} + a^2_{is}} \ \ 
(i = 1, 2, 3, 4) \ .
\label{eq:semiAmp}
\end{equation}
%------------------------------------------------------------

%-----------------------------------------------------------
\subsection{Quality of the four-harmonic fit}
\label{subsection:quality}
%-----------------------------------------------------------

The uniform analysis of a complete sample of more than $10^4$ periodic variables allowed us to examine in details the quality of our model and in particular minute periodic modulations which were not accounted for by the four-harmonic model. To do that we derived the Fourier-transform based power spectrum of the {\it residuals} of each short-period ellipsoidal and identified its 5 highest peaks. These peaks reflected small {\it periodic} modulations, of the order of a few milli-mags, a factor of $10$ or smaller than the ellipsoidal modulation. Nevertheless, in light curves with many measurements and low observational noise, some peaks in the residual periodogram showed high significance. 

\begin{figure} 
\centering
{\includegraphics[scale=0.75]{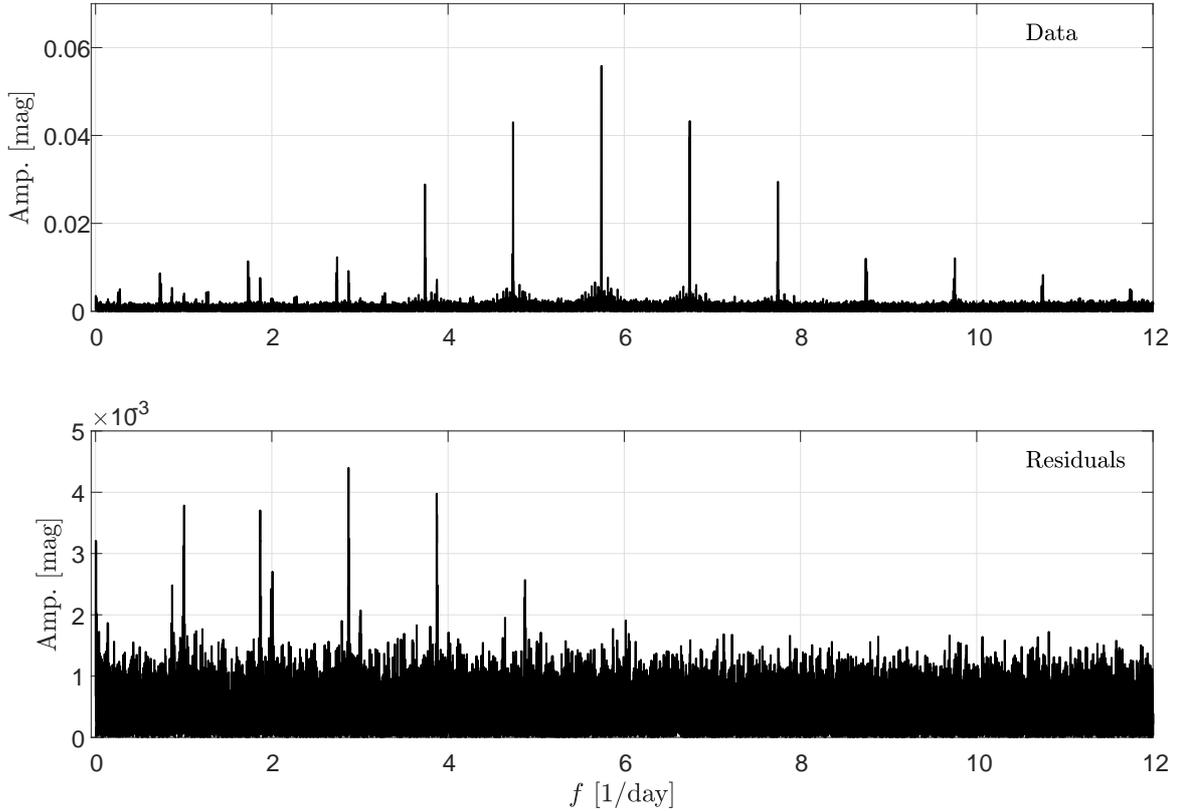}}
\caption{Top --- Magnitude spectrum of OGLE BLG-ELL-018562 light curve, with an orbital period of $\sim 0.348$ day. The spectrum displays a peak of 0.06 mag at $\sim 5.74$ 1/day, corresponding to the dominant second harmonic of the modulation. Other peaks, which appear symmetrically on both sides of the main peak, are due to strong daily aliases induced by the data window function. The first harmonic peak, corresponding to a frequency of $\sim 2.873$ 1/day, is noticeable. 
Bottom --- Magnitude spectrum of residuals. The highest peak corresponds to some power left near the orbital frequency. Other peaks, which appear symmetrically on both sides of the main peak, are due to the data window function. There are also peaks close to integer frequencies, with amplitudes of $2$--$3$ milli-mags. The leftmost peak of $\sim0.003$ mag at $\sim0.0027$ 1/day corresponds to one-year frequency.
}
\label{fig:FFT_018562}
\end{figure}

We considered a frequency to be significant if its corresponding peak was more than $6\sigma$ above the median of the entire power spectrum, with $\sigma$ = $1.48\times$MAD. 
Fig.~\ref{fig:FFT_018562} demonstrates our analysis for OGLE BLG-ELL-018562, in which top 5 frequencies of the {\it residuals} power spectrum are significant. The upper panel shows the magnitude spectrum of the detrended light curve, obtained after subtracting each season's average magnitude. The lower panel shows the magnitude spectrum of residuals, obtained after subtracting from the light-curve its fitted model based on equation~(\ref{eq:harmonics}).

%
%-----------
%-----------

The spectrum of the residuals displays peaks corresponding to the orbital frequency and its daily side-lobe frequencies. There are also peaks close to integer frequencies, with amplitudes of $2$--$3$ milli-mags. The leftmost peak of $\sim0.003$ mag at $\sim0.0027$ 1/day corresponds to the one year frequency.

\begin{figure} 
\centering
{\includegraphics[scale=0.75]{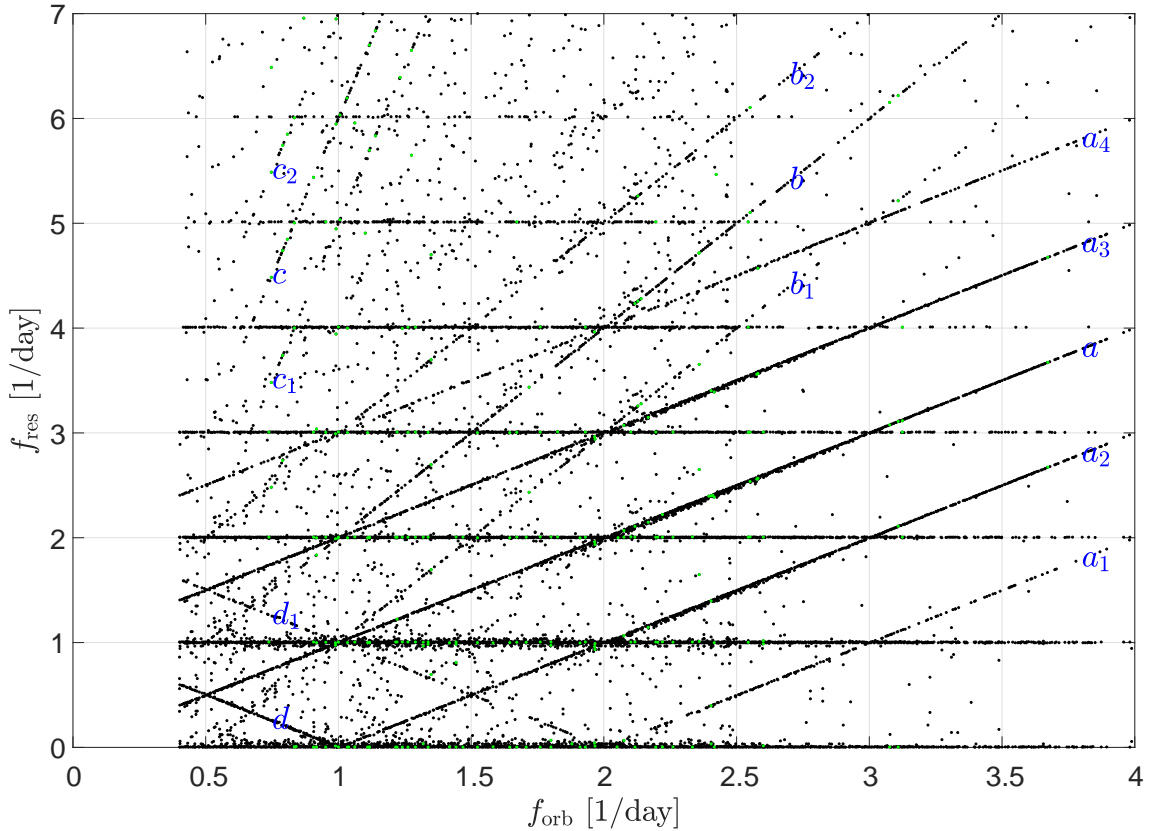}}
\caption{Significant Fourier frequencies of the residuals as a function of the orbital frequency of the OGLE Bulge short-period ellipsoidals. 
The diagram includes up to  five significant frequency peaks found in each power spectrum of the residuals. Only frequencies with $f_{\rm res} \leq 7$ 1/day are shown. The orbital frequency is shown up to $4$ 1/day for clarity.
%
%The vertical lines, which appear close to integer frequencies, are due to the strong daily aliases induced by the window function of the data. The lines with slope of 1 correspond to frequencies of residual side-lobes around the first harmonic, after subtraction of the four-harmonic model. The most left vertical line around $0.0027$ [1/day] corresponds to a sidereal year. 
Candidates with mMMR significantly larger than unity are colored in green.
}
\label{fig:fRES-f}
\end{figure}

We plot in Fig~\ref{fig:fRES-f} $23,744$
significant frequencies that appear in the residuals of all analysed short-period ellipsoidals as a function of the corresponding orbital frequency.

%-----------
%-----------

There are a few clear features in the diagram. 
The concentration of points along integer frequencies are probably due to 
%the strong daily periodicities induced by the daily window function of the data, and by 
imperfect cleaning of daily modulation hidden in the data.
The lowest horizontal line around $0.0027$ 1/day corresponds to an imperfect removal of the sidereal year periodicity.
Again, these periodic modulations are of small amplitudes, on the order of a few milli-mags.

The concentration around the lines denoted by $a$ and $b$ in the figure, with slopes of $1$ or $2$,  respectively, correspond to some power left in the residuals near the orbital frequency and its first harmonic. Parallel to these two lines appear their side-lobe frequencies (see example in Fig.~\ref{fig:FFT_018562}), separated by
$\Delta f=\pm1,\pm2$ relative to the $1$-slope line, and 
$\Delta f=\pm1$ relative to the $2$-slope line.
One can even notice a slight concentration around the slope of $6$, denoted by $c$, corresponding to the sixth-harmonic modulation, which was not part of the fitted model. 

The two lines, denoted by $d$, with a negative slope of $-1$, represent an aliasing of the orbital frequency with the daily window-function frequency. This is relevant only for ellipsoidals with low orbital frequencies.

%==========================================================
\section{Binary Candidates with Probable 
Compact-Object Companions}
\label{sec:candidates}
%==========================================================

%
We now get to the main goal of the paper --- identifying ellipsoidal candidates that might have compact companions. We use equation (1) of \hyperlink{cite.gomel21b}{Paper II} that estimates the ellipsoidal leading amplitude $A_2$, as a function of the fillout factor $f$, the inclination $i$, and the mass ratio $q$: 
%
%---------------------------------------------------------------
%Eq 1.
\begin{equation}
A_2 \approx
\frac{1}{\overline{L}/L_{_0}}\alpha_\mathrm{2} \
f^3 E^3(q) \ q \ \sin^2 i \ C(q,f) \, ,
\label{eq:qminEq}
\end{equation}
%--------------------------------------------------------------
%
where 
$\overline{L}$ is the average luminosity of the star, $L_{_0}$ being the stellar brightness with no secondary at all, and
$E(q)$ is the \cite{eggleton83} approximation for the volume-averaged Roche-lobe radius in binary semi-major axis units.
The ellipsoidal coefficient $\alpha_2$ depends on the linear limb- and gravity-darkening coefficients of the primary and is expected to be in the $1$--$2$ range.
The correction coefficient $C(q,f)$ 
(\hyperlink{cite.gomel21a}{Paper I})
starts at $1$ for $f = 0$ (no correction), as expected, and rises monotonically as $f\to 1$, obtaining a value of $\sim$ 1.5 at $f \gtrsim 0.9$. 
%(for more details see \hyperlink{cite.gomel21b}{Paper II}).

Assuming a fillout factor of $f = 0.98$, inclination of $90^{\circ}$ and a typical $\alpha_2$ of 1.2 for the $I$-band \citep{claret11}, we solve for each ellipsoidal modified minimum mass ratio, mMMR --- $\hat{q}_{\rm min}$, 
based on the observed second harmonic amplitude, $A_2$.
%
%Somewhat arbitrarily, we choose the maximum allowed $\hat{q}_{\rm min}$ to be $20$.
%
The derivation of $\hat{q}_{\rm min}$ assumes that the modulation is due to the ellipsoidal effect of a binary system, and the orbit is circular.

The  uncertainty of $\hat{q}_{\rm min}$ is  inherently large because of the uncertainty of $\alpha_2$, which we somewhat arbitrarily adopt to be $0.1$. Because the resulting distribution of $\hat{q}_{\rm min}$ is highly asymmetric,
we also derive 
\begin{equation}
\hat{q}_{\rm min}^{-1\sigma} \equiv \hat{q}_{\rm min}-\sigma_-(\hat{q}_{\rm min}) \ ,
\end{equation}
to represent the $15.9$ precentile of its distribution,
where 
$\sigma_-(\hat{q}_{\rm min})$
is the one-side $1\sigma$ uncertainty of $\hat{q}_{\rm min}$. 

We excluded two systems from our analysis: OGLE BLG-ELL-015352 and OGLE BLG-ELL-024252. Both systems present a large amplitude of $A_2 \sim 0.25$, for which equation~(\ref{eq:qminEq}) cannot be solved for $\hat{q}_{\rm min}$, within our assumptions. Moreover, all four harmonic coefficients (except $a_{\rm 2c}$) of OGLE BLG-ELL-015352 are less than $5\sigma$ significant, which could imply a different variability type, rather than ellipsoidal, with an actual period that is half the value derived by OGLE. The second system, OGLE BLG-ELL-024252, has an untypical light-curve shape for ellipsoidal variables, with a difference between the maxima of $\sim 0.3$ mag. Omitting these systems, we are left with $10,812$ short-period ellipsoidal variables.

Finally, we chose $136$ ellipsoidals with $\hat{q}_{\rm min}^{-1\sigma}>1$, which we considered as having mMMR significantly larger than unity, and therefore candidates for having compact companions. These systems appear in green in previous figures. Three of these candidates showed $A_2$ slightly larger than the maximum value for which equation~(\ref{eq:qminEq}) has a solution within $\hat{q}_{\rm min} < 20$. An acceptable modification of $\alpha_2$ by less than $25\%$ would provide a solution for $\hat{q}_{\rm min}$, and therefore we assigned $\hat{q}_{\rm min}$ of $20$ to these systems. 

The results of the analysis of all $10,812$ short-period ellipsoidals are given in Table~\ref{tab:Data}. The full table is given as online supplementary material. The table includes the OGLE ID, OGLE orbital period [days], reference time $T_0$, reference-time error, fitted I magnitude (weighted average of the averaged magnitudes, fitted for each season separately), fitted I magnitude error, cosine and sine Fourier coefficients, each followed by its error [mag], total number of analyzed points, total number of analyzed seasons, light-curve time range [days], reduced $\chi^2$ of the four-harmonic model, $\hat{q}_{\rm min}$ and $\hat{q}_{\rm min}^{-1\sigma}$.
Folded light curves and a summary table for the first $15$ compact-secondary candidates are presented  
in Appendix~\ref{app:A} and \ref{app:B}.
Similar figures and a table for all $136$ candidates are given in the online supplementary document.

\begin{figure} 
\centering
{  \includegraphics[scale=0.7]{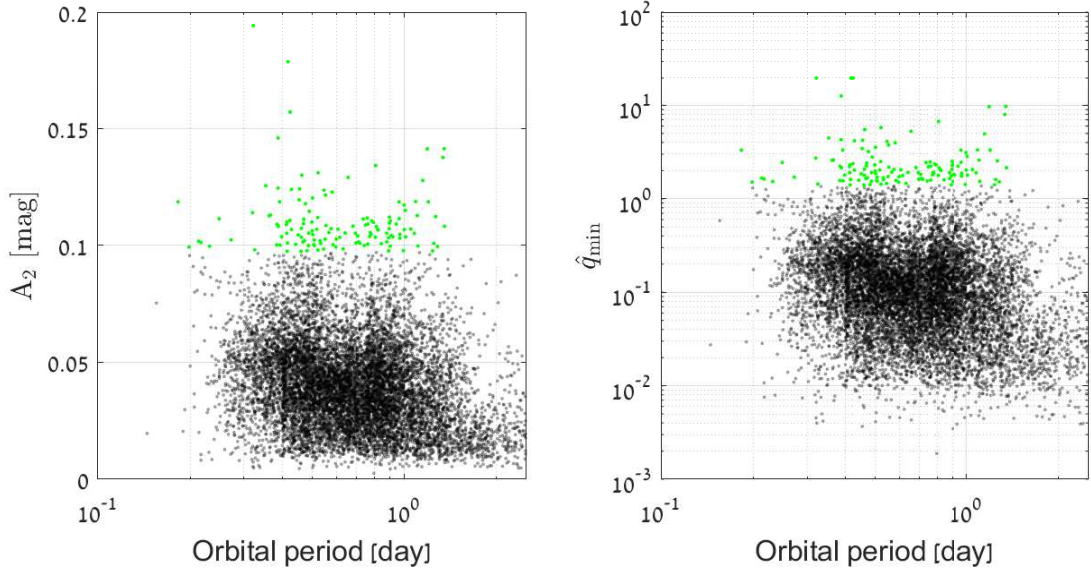}  }
\caption{Second-harmonic amplitude (left) and mMMR (right) as a function of orbital period for the OGLE Bulge short-period ellipsoidals. The mMMR was derived using a typical $\alpha_2$ of $1.2$ for the $I$-band and assuming fillout factor of 0.98. Candidates with mMMR significantly larger than unity are colored in green. 
%The lower envelope visible in the right panel is due to the fact that systems with longer periods have more massive primaries, and therefore a larger range of possible companions, which results in a broader range of available $q$ [MP]. 
}
\label{fig:P-A2-qmin}
\end{figure}

\begin{figure} 
\centering
{  \includegraphics[scale=0.75]{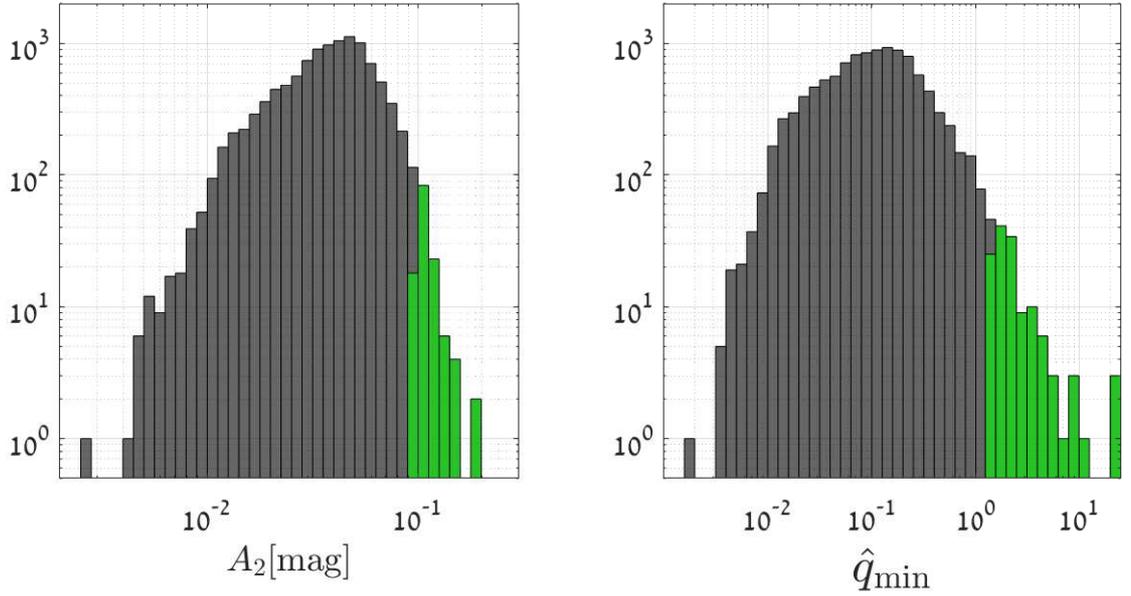}  }
\caption{Second-harmonic amplitude (left) and resulting modified minimum mass-ratio (right) distribution of the OGLE Bulge short-period ellipsoidals, plotted with equally-spaced bins of $\Delta \log A_2 = 0.05$ and $\Delta \log q = 0.1$ correspondingly. Candidates with mMMR significantly larger than unity are colored in green.
}
\label{fig:A2-q-hist}
\end{figure}

The $A_2$ amplitude and the derived $\hat{q}_{\rm min}$ are plotted as a function of the orbital period in Fig.~\ref{fig:P-A2-qmin}, including the $136$ candidates binaries. Their histograms are plotted in Fig.~\ref{fig:A2-q-hist}.

%-----------
%-----------

%-----------
%-----------
 
%

\newpage
%--------------------------------------
\subsection{One example --- OGLE BLG-ELL-024717}
\label{024717}
%--------------------------------------

As an example we present one of our compact-companion candidates, OGLE BLG-ELL-024717, a presumed binary with an orbital period of $P = 0.962$ day. According to our analysis, the $I$-band ellipsoidal modulation presents a relatively large amplitude of $A_2$ = $0.11866 \pm 0.00063$ mag, resulting in  
$\hat{q}_{\rm min}^{-1\sigma} \sim 2.2$.

\begin{figure} 
\centering
{\includegraphics[scale=0.8]{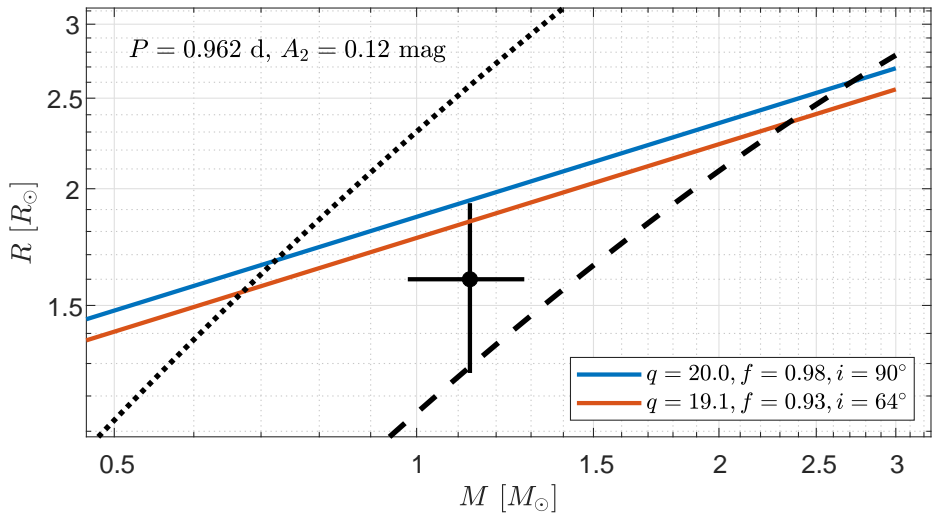}  }
\caption{ Mass-radius relations of OGLE BLG-ELL-024717, derived for $P = 0.962$ d, $A_2 = 0.11866$ mag, and a typical $\alpha_2$ value of $1.2$ for the I band. Blue and red lines correspond to the highest and lowest slopes of the mass-radius relations respectively, obtained for mass ratio, fillout factor and inclination indicated in the legend. Dashed (dotted) line presents $R=R_{\scalebox{0.6}{ZAMS}}$ ($R=2R_{\scalebox{0.6}{ZAMS}}$) from \citet{eker18}, smoothed by a fifth-degree polynomial for clarity. The black point and error bars mark the primary mass and radius and their uncertainties, derived by  \texttt{ISOCHRONES}.
}
\label{fig:OGLE024717}
\end{figure}

As shown in \hyperlink{cite.gomel21b}{Paper II}, one can obtain interesting constraints on the mass-radius relation of the primary star by using \cite{eggleton83} approximation for the volume-averaged Roche-lobe radius and Kepler's third law. Using $P$, $A_2$, and equations~(1) \& (3) of \hyperlink{cite.gomel21b}{Paper II}, we  derive two mass-radius relations for this system, shown in Fig.~\ref{fig:OGLE024717}. 
In the same plot we present mass-radius relations for zero-age main-sequence stars, $R=R_{\scalebox{0.6}{\rm ZAMS}}$, and for $R = 2R_{\scalebox{0.6}{\rm ZAMS}}$. As explained in \hyperlink{cite.gomel21b}{Paper II}, we  expect the primary star of the ellipsoidal binary to be, in most cases, within the region bounded by the four graphs.

To estimate the primary mass and radius of the primary, we used the python \texttt{ISOCHRONES}\footnote{https://isochrones.readthedocs.io/en/latest/} software package \citep{morton15},
used in many studies
\citep[e.g.][]{huber16,mathur17,koposov20}. \texttt{ISOCHRONES}
provides a simple interface to grids of stellar evolution models, and allows determination of stellar properties, such as effective temperature, mass, and radius, based on photometric observations. 

We used the observed photometric magnitudes of GAIA three bands: 
${\rm BP} = 16.942 \pm 0.031$, 
${\rm G} = 16.534 \pm 0.020$, 
${\rm RP} = 15.855 \pm 0.020$, 
adopting a minimum error of $0.02$ mag, and a parallax of $0.345 \pm 0.082$ mas, 
as reported in GAIA EDR3 archival database.\footnote{https://gea.esac.esa.int/archive/}
We assumed a Gaussian reddening prior centered at the TIC\footnote{https://tess.mit.edu/science/tess-input-catalogue/} 
\citep{TIC} value, $E(B-V) = 0.241 \pm0.100$,
adopting a more realistic uncertainty of 0.1 mag. By using the Mesa Isochrones and Stellar Tracks
(MIST) models \citep{choi16} to fit the observed data, we determined an effective temperature of $T_{\rm eff} = 6000 \pm 350 K$, stellar mass of $M_1 = 1.13 \pm 0.16 M_{\odot}$ and radius of $R_1 = 1.60 \pm 0.33 R_{\odot}$ for the primary star, assuming no contribution of light from the secondary.

The location of the primary is presented on the mass-radius diagram using its derived mass and radius and their uncertainties. 
The  uncertainty of the stellar radius reaches our permitted region. The diagram suggests that the star is close to filling its Roche lobe, and its radius is about $1.5$ times larger than the corresponding ZAMS value. The relatively large radius error does not allow us to derive a more accurate minimum mass ratio based on the orbital period, second-harmonic amplitude, and primary mass and radius, as described in \hyperlink{cite.gomel21b}{Paper II}. However, the combination of $M_1$ and $\hat{q}_{\rm min}$ suggests an unseen secondary that might have a mass of $\sim 3.7 M_{\odot}$ or higher. 
Therefore, the companion of OGLE BLG-ELL-024717 might be a black hole.

%
%------------------------------------------------------------------
%---------------------------------------------------------------

%================================
\section{Summary and Discussion}
\label{sec:discussion}
%=================================

We have identified $136$ OGLE Bulge short-period ellipsoidals that might have compact companions, out of $10,956$ short-period ellipsoidals \citep{soszy16} found by the OGLE project \citep{udalski15}. The short-period sample is a subset of the OGLE collection of $450,000$ eclipsing and ellipsoidal binary systems, identified among $\sim 400$ million observed stars towards the Galactic bulge.

The compact-object candidates' selection relies on the identification of these systems as ellipsoidal variables by the OGLE team and the assumption that the binary orbit is circular. 
Obviously, these systems necessitate follow-up radial-velocity observations to establish their binarity and confirm the high mass ratio. This can be done, for example, with the upcoming multi-object spectrographs like VIMOS \citep{garilli14}; FMOS \citep{kimura10} and 4MOST \citep{deJong19},
as was done, for example, 
by \citet{tal-or15} with the  AAOmega multi-object spectrograph \citep{lewis02,smith04}
in the case of CoRoT ellipsoidals \citep{mazeh10,faigler11}.
%At least one should follow-up a random chosen systems from our sample in order to estimate the purity of the sample. 
%
In fact, the 
Gaia\footnote{https://www.cosmos.esa.int/web/gaia/release/\label{fn_Gaia}} mission is planned to release a large number of stellar spectroscopic measurements of our bright candidates, enabling the confirmation of the massive-secondary conjecture. 

Note that the present approach is not meant to be complete. Our analysis misses all binaries with compact-object secondaries for which the primaries are far from filling their Roche lobe. To identify such binaries one needs a reliable mass and radius estimation, and even then binaries with small orbital inclination might be missed. In the OGLE Bulge ellipsoidals there are only $195$ short-period MS binaries with accurate enough distance and extinction derived by Gaia EDR3 \citep{brown20,gaiaEdr3}. These numbers are expected to grow substantially when Gaia DR3\footref{fn_Gaia} 
becomes available. In any case, the ellipsoidal-effect approach is biased towards binaries with large primaries, as the effect depends on the stellar radius to the third power. 

The present approach has high potential, thanks to existing --- e.g., TESS \citep{ricker15} and ASAS-SN \citep[e.g.,][]{kochanek17},
and expected large photometric surveys in the near future, such as LSST \citep[e.g.,][]{abell09,ivezic19} and ZTF \citep{bellm19}. One of the largest photometric survey is obviously that of Gaia, which obtains tens of photometric observations for more than one billion stars \citep{evans18}. Gaia DR3 is expected to identify a large number of ellipsoidal variables and might even publish compact-object candidates.  

Another possible source of dormant BHs is the upcoming Gaia RVS survey \citep{katz19}, planned to be included in DR3. This survey will identify a large sample of spectroscopic binaries by multiple observations of many stars. Obviously, a few of those binaries might have dormant compact-object secondaries. 

In Fig.~\ref{fig:P-WI} we plotted the period-luminosity relation of the Bulge ellipsoidals.
The PL relation formed by red giants was discovered by \citet{wood1999} and was further investigated by \citet{soszynski2004}.
The PL formed by RC stars, lying on the prolongation of the red giant's relation towards the shorter period range was studied by \citet{pawlak2014}. MS ellipsoidals form a PL relation that differs significantly from RC and RGB stars. This relation has been analysed by \citet{rucinski1994, rucinski2006} and \citet{mateo2017} for low-mass stars and by \citet{pawlak2016} for high-mass stars.
\citet{jayasinghe2020} shows that the PL relation for MS stars is different above and below the Kraft break \citep{kraft67}. Fig.~\ref{fig:P-WI} shows that the period cut at 2.5~days separates well the PL relations formed by MS and RC stars, making it a simple robust way to extract the sample of MS stars, which is the focus of this paper.

The PL relation in ellipsoidal binaries is probably because the primary component in these systems is close to filling its Roche lobe. Otherwise, the ellipsoidal amplitude would be too small and not observable. Therefore,  the primary star radius in these ellipsoidals is comparable to the semi-major axis of the orbit, which is correlated with the orbital period by Kepler's third law. However, a thorough study of the PL relation is out of the scope of the present paper and is deferred to an upcoming publication (Pawlak et al., in preparation).

In the next stage of this project we plan to consider the 1159 LMC and 316 SMC OGLE ellipsoidals \citep{pawlak16}. 
The advantage of that sample is the well-known distances of 50 kpc \citep{pietrzynski13} and 62 kpc \citep{graczyk13} 
to the LMC and SMC, respectively, and the small extinction \citep{gordon03,inno16,joshi19}. These features might allow us to estimate the mass and radius of MS stars and therefore derive not only their mMMR but the minimum mass ratio itself. On the other hand, the LMC and SMC samples are painfully small. 

Note that 
late-type MS stars of these two satellite galaxies are too faint for the OGLE project. Instead, OGLE LMC and SMC MS ellipsoidals have early-type stars as their primaries. Therefore, extending this project to the OGLE LMC and SMC ellipsoidals might enable us to search BH companions of early-type stars.
Note that if the primary is an early-type star, of, say, 
$M \gtrsim 3 M_{\odot}$, a system with a mass ratio larger than unity must have a BH component and cannot have an NS secondary. The present study of the Bulge ellipsoidals, on the other hand, might reveal NS or even in some cases white-dwarf secondaries in binaries with late-type MS primaries.

When the two OGLE-based projects will be completed with DR3 information incorporated, and with follow-up RVs obtained, we will be able to better understand the frequency of compact objects in short-period binaries with MS primaries, and their mass distribution in particular \citep{ozel10}. 

Short-period binaries with high-mass primaries,  of say, $M \gtrsim 8 M_{\odot}$, may go through a SuperNova explosion and form a binary with two compact objects. Such a binary might be the progenitor of a Gravitational-Wave burst when the two objects merge.  
We then might even be able to compare the observed rate of compact-object mergers \citep[e.g.,][]{abbott16,abbott17,abbott19} with the progenitor population of short-period binaries with a compact secondary.

%========================
\section*{Acknowledgments}
%========================
%
This research was supported by Grant No.~2016069 of the United States-Israel Binational Science Foundation (BSF) and by the Grant No.~I-1498-303.7/2019 of the
German-Israeli Foundation. 
%
%========================
\section*{Data availability}
%========================
%
The data underlying this article are available in the article and in its online supplementary material. The full table of analysis results is available at the CDS.
%
%================================================
%

%===========================
\bibliographystyle{mnras}
\bibliography{BHIII_bib}
%
%%%%%%%%%%%%%%%%% APPENDICES %%%%%%%%%%%%%%%%%%%%%
%==============================================
%
%
%
\appendix
\section{OGLE light curves}
\label{app:A}

%-----------
\begin{figure} 
\hspace*{-2.2cm}
%\centering
%\resizebox{16cm}{20cm}
{  \includegraphics{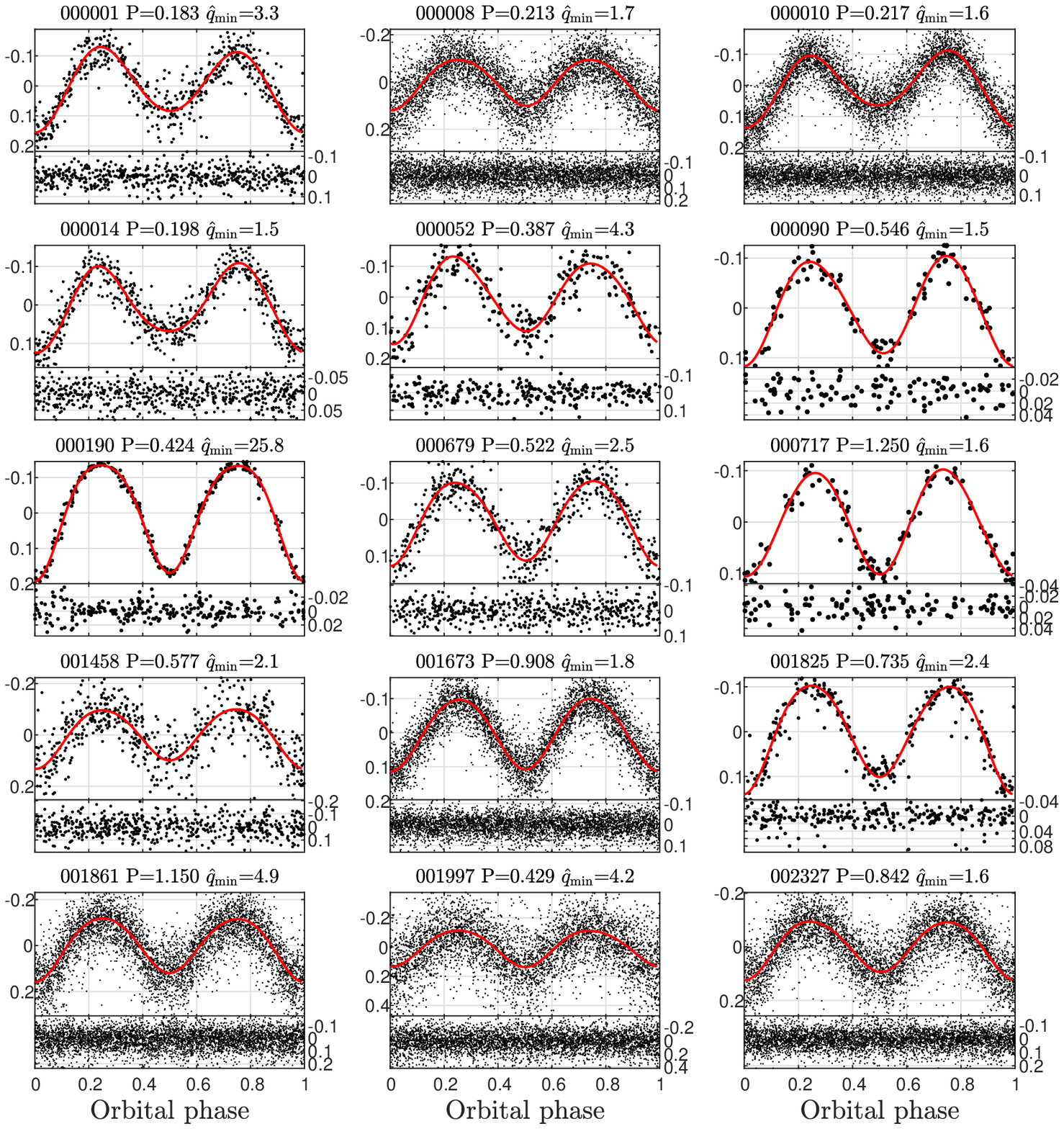}  }
\caption{ Folded detrended OGLE light curves in the $I$ band of the first 15 candidates. The last six digits of the OGLE identifier are indicated together with the OGLE period in days and $\hat{q}_{\rm min}$. Zero phase is defined at the epoch of  the second-harmonic minimum. A four harmonics model is plotted with a solid line. The dominant $A_2$ coefficient gives the characteristic double-peaked appearance to the light curve, while $A_1$ and $A_3$ are related to the difference between the minima. The residuals are plotted in the lower panels. The data (in both panels) are shown between the $1$st--$99$th percentiles for clarity.
}
\label{fig:lcs}
\end{figure}
%-----------

\section{Data table}
\label{app:B}

\newcommand*{\thead}[1]{\multicolumn{1}{|c|}{\bfseries #1}}
\renewcommand{\arraystretch}{2}
\begin{table*}
    \Rotatebox{90}{
    \resizebox{\textwidth}{!}{
	\begin{tabular}{|r|r|r|r|r|r|r|r|r|r|r|r|r|r|r|r|r|r|}
		\hline
		 \thead{ID} & \thead{$P$} & \thead{$T_0$} & \thead{$I$} & \thead{$a_{\rm 1c}$} & \thead{$a_{\rm 2c}$} & \thead{$a_{\rm 3c}$} & \thead{$a_{\rm 4c}$} & \thead{$a_{\rm 1s}$} & \thead{$a_{\rm 2s}$} & \thead{$a_{\rm 3s}$} & \thead{$a_{\rm 4s}$} & \thead{$N$} & \thead{$N_{\rm s}$} & \thead{$\Delta t$} & \thead{$\chi^2_{\rm red}$} & \thead{$\hat{q}_{\rm min}$} & \thead{$\hat{q}_{\rm min}^{-1\sigma}$} \\
		 \thead{} & \thead{} & \thead{$T_{\rm 0,err}$} & \thead{$I_{\rm err}$} & \thead{$a_{\rm 1c,err}$} & \thead{$a_{\rm 2c,err}$} & \thead{$a_{\rm 3c,err}$} & \thead{$a_{\rm 4c,err}$} & \thead{$a_{\rm 1s,err}$} & \thead{$a_{\rm 2s,err}$} & \thead{$a_{\rm 3s,err}$} & \thead{$a_{\rm 4s,err}$} & \thead{} & \thead{} & \thead{} & \thead{} & \thead{} & \thead{} \\
		 \hline
        000001 & 0.182739 & 6171.72930 & 18.4002 & -0.0235 & 0.1187 & -0.0111 & 0.0005 & 0.0060 & 0.0000 & -0.0017 & -0.0034 & 579 & 6 & 2029 & 2.5 & 3.3 & 2.2\\
        &  & 0.00037 & 0.0018 & 0.0026 & 0.0026 & 0.0026 & 0.0026 & 0.0024 & 0.0025 & 0.0025 & 0.0025 &  &  &  &  &  & \\
        \hline 
        000008 & 0.213058 & 6393.89664 & 19.0265 & -0.0095 & 0.1019 & -0.0001 & -0.0090 & 0.0016 & 0.0000 & 0.0015 & -0.0021 & 4995 & 6 & 2072 & 2.8 & 1.7 & 1.2\\
        &  & 0.00025 & 0.0010 & 0.0015 & 0.0015 & 0.0014 & 0.0014 & 0.0014 & 0.0014 & 0.0014 & 0.0014 &  &  &  &  &  & \\
        \hline 
        000010 & 0.217000 & 6392.91249 & 18.18486 & -0.02337 & 0.10130 & -0.01197 & 0.00212 & -0.01026 & 0.00000 & -0.00182 & -0.00203 & 4972 & 6 & 2072 & 3.7 & 1.6 & 1.2\\
        &  & 0.00019 & 0.00066 & 0.00099 & 0.00094 & 0.00094 & 0.00095 & 0.00090 & 0.00096 & 0.00094 & 0.00094 &  &  &  &  &  & \\
        \hline 
        000014 & 0.198242 & 6117.20526 & 17.8945 & -0.0128 & 0.0994 & -0.0148 & 0.0036 & -0.0056 & 0.0000 & -0.0015 & -0.0022 & 646 & 6 & 2069 & 3.1 & 1.5 & 1.1\\
        &  & 0.00040 & 0.0013 & 0.0019 & 0.0019 & 0.0019 & 0.0019 & 0.0018 & 0.0019 & 0.0019 & 0.0019 &  &  &  &  &  & \\
        \hline 
        000052 & 0.387288 & 5790.6487 & 18.2595 & -0.0095 & 0.1246 & -0.0110 & -0.0066 & 0.0019 & 0.0000 & -0.0066 & -0.0084 & 271 & 4 & 1311 & 2.4 & 4.3 & 2.7\\
        &  & 0.0014 & 0.0028 & 0.0042 & 0.0042 & 0.0038 & 0.0039 & 0.0039 & 0.0039 & 0.0042 & 0.0042 &  &  &  &  &  & \\
        \hline 
        000090 & 0.545850 & 2729.8142 & 17.7955 & -0.0094 & 0.0985 & -0.0035 & -0.0021 & 0.0014 & 0.0000 & 0.0073 & -0.0038 & 100 & 2 & 593 & 1.2 & 1.5 & 1.0\\
        &  & 0.0019 & 0.0021 & 0.0032 & 0.0031 & 0.0030 & 0.0029 & 0.0030 & 0.0031 & 0.0031 & 0.0034 &  &  &  &  &  & \\
        \hline 
        000190 & 0.424034 & 5788.36752 & 14.72946 & -0.0071 & 0.1571 & -0.0046 & -0.0228 & 0.0004 & 0.0000 & -0.0007 & 0.0001 & 270 & 4 & 1311 & 22 & 26 & 11\\
        &  & 0.00038 & 0.00097 & 0.0014 & 0.0014 & 0.0014 & 0.0014 & 0.0014 & 0.0014 & 0.0014 & 0.0014 &  &  &  &  &  & \\
        \hline 
        000679 & 0.522313 & 6169.7456 & 18.1462 & -0.0022 & 0.1120 & -0.0049 & -0.0088 & -0.0003 & 0.0000 & 0.0023 & -0.0011 & 581 & 6 & 2029 & 3.4 & 2.5 & 1.7\\
        &  & 0.0011 & 0.0016 & 0.0024 & 0.0023 & 0.0023 & 0.0023 & 0.0022 & 0.0023 & 0.0024 & 0.0023 &  &  &  &  &  & \\
        \hline 
        000717 & 1.249635 & 2555.5971 & 17.4605 & -0.0085 & 0.1009 & 0.0078 & -0.0038 & -0.0032 & 0.0000 & 0.0000 & -0.0021 & 98 & 2 & 598 & 1.9 & 1.6 & 1.1\\
        &  & 0.0032 & 0.0018 & 0.0025 & 0.0025 & 0.0026 & 0.0026 & 0.0027 & 0.0026 & 0.0026 & 0.0025 &  &  &  &  &  & \\
        \hline 
        001458 & 0.577088 & 6178.3318 & 18.8606 & -0.0135 & 0.1073 & -0.0033 & -0.0087 & -0.0026 & 0.0000 & -0.0009 & -0.0013 & 576 & 6 & 2047 & 2.7 & 2.1 & 1.4\\
        &  & 0.0025 & 0.0029 & 0.0043 & 0.0041 & 0.0041 & 0.0040 & 0.0039 & 0.0041 & 0.0041 & 0.0043 &  &  &  &  &  & \\
        \hline 
        001673 & 0.908019 & 6220.92755 & 18.50546 & -0.00575 & 0.10429 & 0.00268 & -0.00660 & -0.00047 & 0.00000 & 0.00092 & -0.00036 & 4602 & 6 & 2055 & 2.5 & 1.8 & 1.3\\
        &  & 0.00073 & 0.00062 & 0.00091 & 0.00088 & 0.00087 & 0.00088 & 0.00084 & 0.00088 & 0.00088 & 0.00087 &  &  &  &  &  & \\
        \hline 
        001825 & 0.734745 & 6457.9357 & 16.2996 & -0.0115 & 0.1114 & -0.0077 & -0.0109 & 0.0023 & 0.0000 & 0.0007 & 0.0016 & 286 & 5 & 1671 & 35 & 2.4 & 1.7\\
        &  & 0.0020 & 0.0019 & 0.0028 & 0.0027 & 0.0026 & 0.0027 & 0.0026 & 0.0027 & 0.0027 & 0.0027 &  &  &  &  &  & \\
        \hline 
        001861 & 1.149962 & 6377.9891 & 19.1131 & -0.0144 & 0.1278 & -0.0053 & -0.0115 & 0.0014 & 0.0000 & -0.0003 & 0.0012 & 4800 & 6 & 2055 & 2.6 & 4.9 & 3.1\\
        &  & 0.0014 & 0.0010 & 0.0015 & 0.0015 & 0.0014 & 0.0014 & 0.0014 & 0.0014 & 0.0014 & 0.0014 &  &  &  &  &  & \\
        \hline 
        001997 & 0.428522 & 6427.0104 & 19.7377 & -0.0030 & 0.1240 & 0.0042 & -0.0112 & -0.0006 & 0.0000 & -0.0019 & -0.0034 & 4286 & 6 & 2052 & 3.0 & 4.2 & 2.6\\
        &  & 0.0011 & 0.0020 & 0.0030 & 0.0029 & 0.0028 & 0.0028 & 0.0027 & 0.0028 & 0.0028 & 0.0028 &  &  &  &  &  & \\
        \hline 
        002327 & 0.842062 & 6378.93513 & 18.81913 & -0.0095 & 0.1008 & -0.0057 & -0.0083 & 0.0020 & 0.0000 & 0.0011 & -0.0021 & 4809 & 6 & 2055 & 3.1 & 1.6 & 1.1\\
        &  & 0.00093 & 0.00090 & 0.0013 & 0.0013 & 0.0013 & 0.0013 & 0.0012 & 0.0013 & 0.0013 & 0.0013 &  &  &  &  &  & \\
        \hline 
	\end{tabular}
	}
	}
	\\
	\vspace{3mm}
	\caption{ Fitted parameters of the first 15 candidates: OGLE ID (last six digits), OGLE orbital period [days], reference time [JD-2450000], reference-time error, fitted I magnitude, fitted I magnitude error, cosine and sine Fourier coefficients, each followed by its error [mag], total number of analyzed points, total number of analyzed seasons, light-curve time range [days], reduced $\chi^2$ of the four-harmonic model, mMMR and lower-percentile mMMR. The full table is available at the CDS. 
	}
	\label{tab:Data}
\end{table*} 

%====================
%
% Don't change 
% Don't change these lines
\bsp	% typesetting comment
\label{lastpage}
\end{document}